% TEX SET-UP FOR PROCEEDINGS OF "ASI MEETING HELD AT AHMEDABAD"
% TO BE PUBLISHED IN THE BULLETIN OF ASTRONOMICAL SOCIETY OF INDIA
% Please do not change the page layout, except \hoffset and/or
% \voffset, where changes may be required depending on default printer
% settings. Leave a margin of 3.8 cm on the left and 5 cm at the top
\input psfig.sty
\magnification=\magstep0
\hsize=13.5 cm               %  horizontal size of printed page
\vsize=19.0 cm               %  vertical size of printed page
\baselineskip=12 pt plus 1 pt minus 1 pt  % The line spacing
\parindent=0.5 cm  % The paragraph indent
\hoffset=1.3 cm      % The horizontal offset (may need to be changed)
\voffset=2.5 cm      % The vertical offset (may need to be changed)
\font\twelvebf=cmbx10 at 12truept % Set bold font for Title
\font\twelverm=cmr10 at 12truept % Set large font for Name
\overfullrule=0pt
\nopagenumbers    %  Actual page nos. will be inserted by the Editor
%
% The headlines
% The changes in the headlines should be made just before the Abstract
\newtoks\leftheadline \leftheadline={\hfill {\eightit S. K. Saha and 
V. Chinnappan}
\hfill}
\newtoks\rightheadline \rightheadline={\hfill {\eightit Night time variation
of Fried's parameter}
 \hfill}
% Do not change the headline on the first page of paper.
\newtoks\firstheadline \firstheadline={{\eightrm Bull. Astron. Soc.
India (1998) {\eightbf 26,} } \hfill}
\def\makeheadline{\vbox to 0pt{\vskip -22.5pt
\line{\vbox to 8.5 pt{}\ifnum\pageno=1\the\firstheadline\else%
\ifodd\pageno\the\rightheadline\else%
\the\leftheadline\fi\fi}\vss}\nointerlineskip}
%
% Defining 8-pt fonts for figure captions and references
\font\eightrm=cmr8  \font\eighti=cmmi8  \font\eightsy=cmsy8
\font\eightbf=cmbx8 \font\eighttt=cmtt8 \font\eightit=cmti8
\font\eightsl=cmsl8
\font\sixrm=cmr6    \font\sixi=cmmi6    \font\sixsy=cmsy6
\font\sixbf=cmbx6
%
%for switching to eight point type \eightpoint
\def\eightpoint{\def\rm{\fam0\eightrm}
\textfont0=\eightrm \scriptfont0=\sixrm \scriptscriptfont0=\fiverm
\textfont1=\eighti  \scriptfont1=\sixi  \scriptscriptfont1=\fivei
\textfont2=\eightsy \scriptfont2=\sixsy \scriptscriptfont2=\fivesy
\textfont3=\tenex   \scriptfont3=\tenex \scriptscriptfont3=\tenex
\textfont\itfam=\eightit  \def\it{\fam\itfam\eightit}%
\textfont\slfam=\eightsl  \def\sl{\fam\slfam\eightsl}%
\textfont\ttfam=\eighttt  \def\tt{\fam\ttfam\eighttt}%
\textfont\bffam=\eightbf  \scriptfont\bffam=\sixbf
\scriptscriptfont\bffam=\fivebf \def\bf{\fam\bffam\eightbf}%
\normalbaselineskip=10pt plus 0.1 pt minus 0.1 pt
\normalbaselines
\abovedisplayskip=10pt plus 2.4pt minus 7pt
\belowdisplayskip=10pt plus 2.4pt minus 7pt
\belowdisplayshortskip=5.6pt plus 2.4pt minus 3.2pt \rm}
%
% define the displayed equations to be indented 1.5 cm from left
% as required by the Bulletin. Hopefully this will work for all
% equations. With this definition using the normal $$...$$ should
% produce the equations with correct indentation, but it will be necessary
% to use \eqno to put equation numbers (though it will be possible to put
% blank eq. nos.). Further, eq. nos using \eqalignno will not work
%
\def\leftdisplay#1\eqno#2$${\line{\indent\indent\indent%
$\displaystyle{#1}$\hfil #2}$$}
\everydisplay{\leftdisplay}
%
% Some useful definitions
% less than or order of \la
\def\frac#1#2{{#1\over#2}}

% greater than or order of \ga

%
%
%to generate boldface characters
\def\pmb#1{\setbox0=\hbox{$#1$}\kern-0.015em\copy0\kern-\wd0%
\kern0.03em\copy0\kern-\wd0\kern-0.015em\raise0.03em\box0}
%
%Beginning of Document%
\pageno=1
\vglue 50 pt  %Leave some space on page 1 before the title
% The title
%
\leftline{\twelvebf  Night time variation of Fried's parameter at VBT, Kavalur} 
% if more than one line is required for the title, then use next two lines ...
%
\smallskip
%\leftline{\twelvebf Proceedings of ``ASI meeting held at Ahmedabad''}
% end of title
\vskip 40 pt  % Space between title and author(s) name(s).
\leftline{\twelverm S. K. Saha and V. Chinnappan} % Name of Authors
\vskip 4 pt
\leftline{\eightit Indian Institute of Astrophysics, Bangalore 560 034, India.
}
%\leftline{\eightit  to reduce the number of lines}
%
% If authors are from different institutes, repeat the above lines
% for each institution. For authors from same institution write the
% names in one line.
%
%\vskip 0.5 cm
%\leftline{\twelverm V. R. Co-author1 and V. R. Co-author2}
%\vskip 4 pt
%\leftline{\eightit Name and Address of the institution}
\vskip 20 pt % leave some space between author(s) names(s) and abstract
%
%
% The leftheadline should include the Authors' name, for two authors use
% \&  (e.g. I. M. Author \& I. M. Co-author) for three or more authors
% use et al.,
\leftheadline={\hfill {\eightit S. K. Saha and V. Chinnappan} \hfill}
% Use a short running title as the rightheadline
\rightheadline={\hfill {\eightit Night time variation of Fried's parameter 
at VBT, Kavalur}  \hfill}

% Abstract begins
%
{\parindent=0cm\leftskip=1.5 cm

{\bf Abstract.}
\noindent
The night time variation of the Fried's parameter, r$_{o}$ is measured  
on 28-29 March 1991, at the Cassegrain focus of 2.34 m Vainu Bappu Telescope 
(VBT), at Vainu Bappu Observatory (VBO), Kavalur, India, using speckle 
interferometer; the results are discussed.
\smallskip 
\vskip 0.5 cm  %  Space between Abstract and Key words
{\it Key words:} interferometer, speckle imaging, seeing, Fried's parameter.

}                                 %  End of abstract
% Beginning of document
%
%
% Beginning of a section heading
%
% for the first section leave 20 pt space, for subsequent sections just
% leave bigskip (i.e. 12 pt)
\vskip 20 pt
\centerline{\bf 1. Introduction}
\bigskip
\noindent  
When a wavefront passes down through the atmosphere, 
it suffers phase fluctuations and reaches the entrance pupil of a telescope 
with patches of random excursions in phase (Fried, 1966). If the exposure time 
is shorter ($<$20 ms) than the evolution time of the phase 
inhomogeneities, then each patch of the wavefront with diameter r$_{o}$ would 
act independently of the rest of the wavefront resulting in
many bright spots - speckles - spread over the area defined by the long
exposure image. These speckles can occur randomly along any direction within
an angular patch of diameter 1.22 $\lambda$/r$_{o}$. We present here 
the night time variation of r$_{o}$ from the data obtained at 2.34 m  
VBT, at VBO, Kavalur, using speckle interferometric technique (Labeyrie, 1970).  
\vskip 20 pt
\centerline{\bf 2. Observations and data processing}
\bigskip 
\noindent
We have recorded speckle-grams of 14 unresolved stars (15 data points) 
in and around 30$^{o}$ zenith with the speckle interferometer (Saha et al., 
1987) on 28-29 March 1991, at the Cassegrain focus of the VBT through a 5nm 
filter centered on H$\alpha$ using frame transfer uncooled intensified CCD 
(Chinnappan et al., 1991). The image at the said focus is sampled to 0.027 
arc second per pixel. The m$_v$ of these stars varied from 5 to 7. The 
observations were carried out between 1500 and 2330 hrs. UT. 
\bigskip 
The averaged autocorrelation of the short exposure images
of a point source contains autocorrelation of the seeing disk together with
the autocorrelation of mean speckle cell (Wood, 1985, Saha et al., 1998). 
Figure 1 depicts the autocorrelation of HR2305 observed at 1510 hrs. UT,
comprising the width of the seeing disk, as well as the width of the speckle
component. The size of the r$_o$ is found to be 8,5 cm at H$\alpha$. 
The form of transfer function $<\mid \widehat{P}(\bf r) \mid^{2}>$ is 
obtained by calculating Wiener spectrum of the instantaneous intensity 
distribution from each of these stars. Here, $\widehat{P}$ is the transfer 
function, ${\bf r}$ = (x,y) is a 2-dimensional space vector, 
$< >$ indicates the ensemble average and $\mid \mid$ the modulus. The r$_o$ is
measured from the two 
speckle-grams acquired at an exposure times of 20 ms, containing odd and even 
fields of a single frame. The average of these frame is the instantaneous 
value of r$_o$ at the time of observation. The estimated error in this 
measurement is of the order of $\pm$ 0.05 arcsec. Figure 2 depicts the night 
time variation of r$_{o}$ on 28-29 March 1991 comprising of zenith
distance corrected value (solid line) and uncorrected value.
\bigskip
\noindent
% inserting figures.
\midinsert
%Leave appropriate space to paste the figure
%\vskip 6.0cm
{\eightpoint   % Switch to 8 pt fonts for figure captions
\noindent
\centerline{\psfig{figure=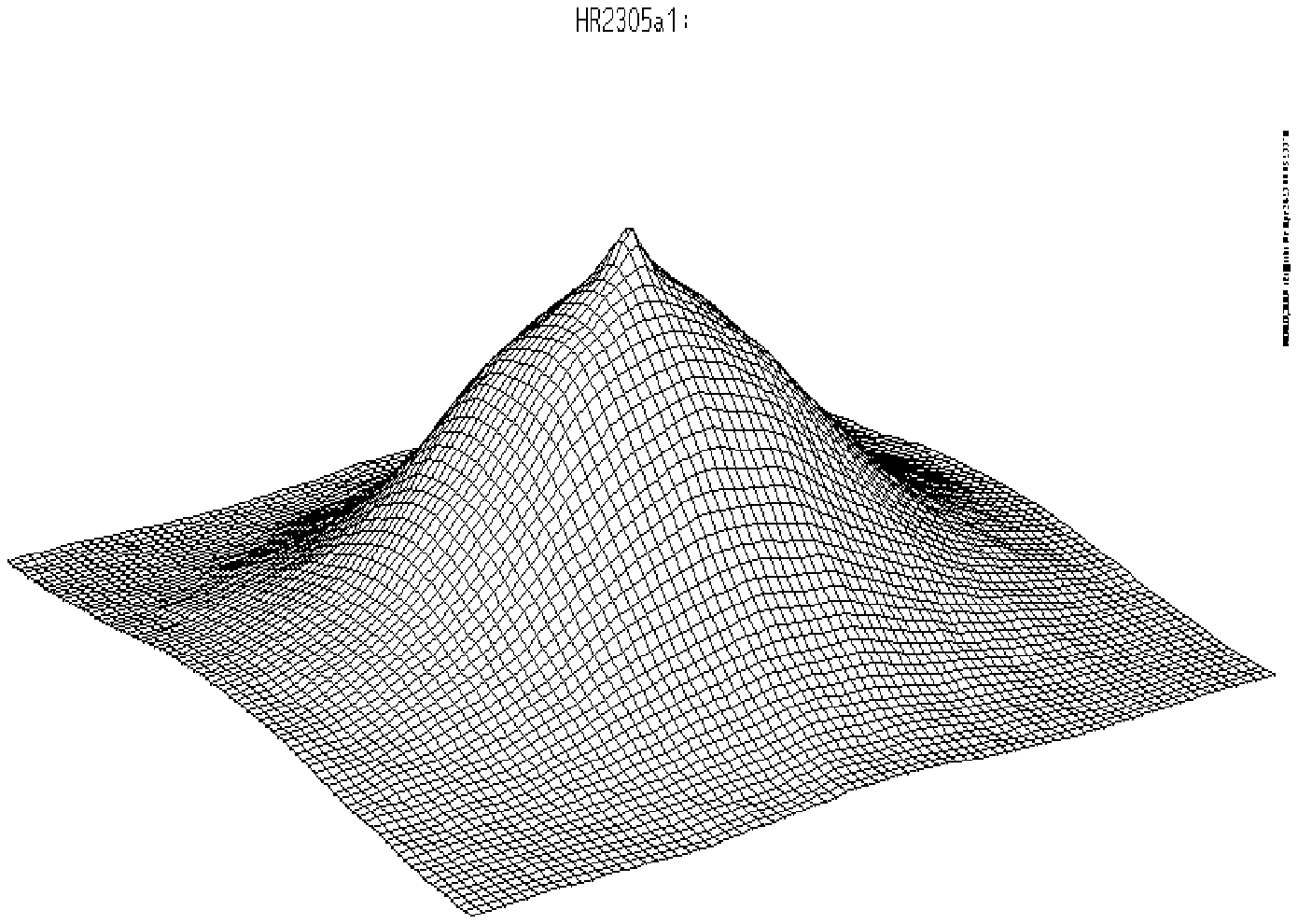,height=6cm,width=7.5cm,angle=270}}
{\bf Figure 1. Autocorrelation of HR2305 observed at 1510 hrs.}  
}
%  End of 8 pt fonts, leave one line blank after the caption
%        to set the caption with reduced baselineskip for 8-pt font.
\endinsert
\noindent
% inserting figures.
\midinsert
%Leave appropriate space to paste the figure
{\eightpoint   % Switch to 8 pt fonts for figure captions
\noindent
\centerline{\psfig{figure=r_o1.ps,height=6cm,width=9cm}}
{\bf Figure 2.} Night time variation of r$_{o}$ on 28-29 Mar., 1991, at 
VBT, Kavalur.
}
%  End of 8 pt fonts, leave one line blank after the caption
%        to set the caption with reduced baselineskip for 8-pt font.
\endinsert
\vskip 20 pt
\centerline{\bf 3. Discussion and Conclusions}
\bigskip 
\noindent
Systematic studies of r$_{o}$ would enable one to understand the various 
causes of the local seeing, for example, thermal inhomogeneities associated 
with the building, aberrations in the design, manufacture and alignment of the 
optical train, etc. The value of r$_{o}$ is an essential parameter in designing 
proper adaptive optics system for the telescope. It is seen from figure 2  
that the seeing improves gradually in the later part of the night at an
interval of several minutes. Racine (1996) too 
observed that the best seeing condition lasts only for several minutes.  
It may be necessary to maintain a uniform temperature in and around
the primary mirror of the telescope to avoid the degradation of the seeing. 
Care has been taken while designing the new speckle interferometer (Saha et 
al., 1997, 1998) to avoid the formation of eddies produced by the hot air 
entrapment.  
\bigskip
Depending upon the high velocity wind, the coherence time varies from $<$1 
ms to $\sim$ 0.1 s. 5$\%$ to 50$\%$ variations in r$_o$ are common within 
a few seconds; they can reach up to 100$\%$ sometimes (Foy, 1988). 
These variations increase the noise in the power spectrum. If the speckle
pattern is not frozen enough due to the long integration time, intermediate
spatial frequencies vanish rapidly. There is loss in signal-to noise ratio and
loss in spatial resolution. The ICCD has fixed integration time, therefore,
a photon counting system of very high time resolution of the order of a few MHz 
(Graves et al., 1993, Papaliolios et al., 1985) is needed to tune the 
integration time according to the value of r$_o$. 
% Sample references
\bigskip
\centerline{\bf References}
\bigskip
{\eightpoint\parindent=0pt\everypar={\hangindent=0.5 cm}
% References in the format of the Bulletin of the Astronomical Society of India
% using 8 pt fonts
% leave one line blank between two references to force a paragraph break
%scussions.

Chinnappan V., Saha S. K., Faseehana, 1991, Kod. Obs. bull. {\bf 11}, 87.
 
Foy R., 1988, Proc. "Instrumentation for Ground-Based Optical Astronomy - 
Present and future" ed. L. Robinson, Springer-Verlag, NY, 345.
 
Fried D. C., 1966, J. Opt. Soc. Am., {\bf 56}, 1972.
 
Graves J. E., Northcott M. J., Roddier C., Roddier F., Anuskiewicz J., Monet G.,
Rigaut F., Madec P. Y., 1993, Proc. ICO-16 Satellite Conf. on "Active and
Adaptive Optics" ed. F. Merkle, Garching bei M$\ddot{u}$nchen, Germany, 47. 
  
Labeyrie A., 1970, A \& A, {\bf 6}, 85.
 
Papaliolios C., Nisenson P., Ebstein S., 1985, App. Opt. {\bf 24}, 287.

Racine R., 1996, P A S P, {\bf 108}, 372.
 
Saha S. K., Jayarajan A. P., Sudheendra G., Umesh Chandra A., 1997, BASI, 
{\bf 25}, 379.

Saha S. K., Sudheendra G., Umesh Chandra A., Chinnappan V., 1998,  
Experimental Astronomy (in press).

Saha S. K., Venkatakrishnan P., Jayarajan A. P., Jayavel N., 1987, Curr. Sci. 
{\bf 57}, 985.

Wood P. R., 1985, Proc. A S A, {\bf 6}, 120. 

% End of section heading
%\endref
}                                         % End of references
% leave one line blank before the closing braces.
\end